\documentclass[conference]{IEEEtran}
\IEEEoverridecommandlockouts
\usepackage{cite}
\usepackage{amsmath,amssymb,amsfonts}
\usepackage{algorithmic}
\usepackage{graphicx}
\usepackage{textcomp}
\usepackage{xcolor}
\usepackage{caption}
\usepackage{subcaption}
\usepackage{multirow}
\usepackage{amsmath}
\usepackage[symbol]{footmisc}

\def\BibTeX{{\rm B\kern-.05em{\sc i\kern-.025em b}\kern-.08em
    T\kern-.1667em\lower.7ex\hbox{E}\kern-.125emX}}

\makeatletter
\def\endthebibliography{%
  \def\@noitemerr{\@latex@warning{Empty `thebibliography' environment}}%
  \endlist
}
\makeatother

\begin{document}

\title{Accelerating Large Language Model Training with Hybrid GPU-based Compression\\
}

\makeatletter
\newcommand{\linebreakand}{%
  \end{@IEEEauthorhalign}
  \hfill\mbox{}\par
  \mbox{}\hfill\begin{@IEEEauthorhalign}
}
\makeatother

\author{
\IEEEauthorblockN{Lang Xu}
\IEEEauthorblockA{\textit{The Ohio State University} \\
Columbus, Ohio \\
xu.3304@osu.edu}
\and
\IEEEauthorblockN{Quentin Anthony}
\IEEEauthorblockA{\textit{The Ohio State University} \\
Columbus, Ohio \\
anthony.301@osu.edu}
\and
\IEEEauthorblockN{Qinghua Zhou}
\IEEEauthorblockA{\textit{The Ohio State University} \\
Columbus, Ohio \\
zhou.2595@osu.edu}
\and
\IEEEauthorblockN{Nawras Alnaasan}
\IEEEauthorblockA{\textit{The Ohio State University} \\
Columbus, Ohio \\
alnaasan.1@osu.edu}
\linebreakand
\IEEEauthorblockN{Radha Gulhane}
\IEEEauthorblockA{\textit{The Ohio State University} \\
Columbus, Ohio \\
gulhane.2@osu.edu}
\and
\IEEEauthorblockN{Aamir Shafi}
\IEEEauthorblockA{\textit{The Ohio State University} \\
Columbus, Ohio \\
shafi.16@osu.edu}
\and
\IEEEauthorblockN{Hari Subramoni}
\IEEEauthorblockA{\textit{The Ohio State University} \\
Columbus, Ohio \\
subramoni.1@osu.edu}
\and
\IEEEauthorblockN{Dhabaleswar K. (DK) Panda}
\IEEEauthorblockA{\textit{The Ohio State University} \\
Columbus, Ohio \\
panda.2@osu.edu}
}
\maketitle

\let\clearpage\relax
\begin{abstract}
    Data Parallelism (DP), Tensor Parallelism (TP), and Pipeline Parallelism (PP) are the three strategies widely adopted to enable fast and efficient Large Language Model (LLM) training. However, these approaches rely on data-intensive communication routines to collect, aggregate, and re-distribute gradients, activations, and other important model information, which pose significant overhead. Co-designed with GPU-based compression libraries, MPI libraries have been proven to reduce message size significantly, and leverage interconnect bandwidth, thus increasing training efficiency while maintaining acceptable accuracy. 
    
    In this work, we investigate the efficacy of compression-assisted MPI collectives under the context of distributed LLM training using 3D parallelism and ZeRO optimizations. We scaled up to 192 V100 GPUs on the Lassen supercomputer. First, we enabled a na\"ive compression scheme across all collectives and observed a 22.5\% increase in TFLOPS per GPU and a 23.6\% increase in samples per second for GPT-NeoX-20B training. Nonetheless, such a strategy ignores the sparsity discrepancy among messages communicated in each parallelism degree, thus introducing more errors and causing degradation in training loss. Therefore, we incorporated hybrid compression settings toward each parallel dimension and adjusted the compression intensity accordingly. Given their low-rank structure \cite{bian2023does}, we apply aggressive compression on gradients when performing DP All-reduce. We adopt milder compression to preserve precision while communicating activations, optimizer states, and model parameters in TP and PP. Using the adjusted hybrid compression scheme, we demonstrate a 17.3\% increase in TFLOPS per GPU and a 12.7\% increase in samples per second while reaching baseline loss convergence. \footnote[1]{This research is supported in part by NSF grants 1818253, 1854828, 2007991, 2018627, 2311830, 2312927, and XRAC grant NCR-130002.}
\end{abstract}

\section{Introduction}
In recent years, an abundance of Large Language Models (LLM) emerged with impressive abilities in downstream Natural Language Processing (NLP) tasks involving machine translation, dialogue systems, text generation, and so on. Some spotlights include LLaMA \cite{touvron2023llama}, GPT-4 \cite{openai2023gpt4} and GPT-NeoX-20B \cite{black2022gptneox20b}. However, to ensure exceptional performance, these models often scale up to billions of parameters, thus requiring increased data size and computation by multiple orders of magnitude. Since the introduction of scaling laws of LLMs \cite{kaplan2020scaling}, model size has been growing from 100 million (BERT \cite{devlin2019bert}) to 500 billion (Megatron-Turing NLG \cite{smith2022using}). As a result, one Graphic Processing Unit (GPU) cannot fit a model and its input data anymore, making it necessary to scale out to more workers. 

High-Performance Computing (HPC) systems are designed and engineered to support sizeable scientific research and deep learning workloads. These HPC systems typically consist of thousands of nodes equipped with two to four advanced GPUs that maximize floating point operations per second (FLOPS), making them ideal for large-scale data-intensive distributed pre-training of LLMs. Inter- and Intra-node communication play a significant role in accelerating parallel applications. The Message Passing Interface (MPI) supports a variety of highly-optimized communication routines and has been a favored parallel programming model deployed on HPC systems \cite{mpich}. With the advances in GPUDirect technology \cite{gpudirect}, GPU-aware MPI libraries \cite{openmpi, mvapich, spectrum-mpi} vastly accelerate GPU data transfer and leverage interconnect bandwidth. MPI also serves as a popular communication back-end for distributed machine learning jobs \cite{gems, hyfi, DP-Super-Resolution}.

Training massive language models requires meticulous arrangement of memory resources and parallelism strategies. Two well-known solutions to this problem are 3D parallelism \cite{smith2022using} and the Zero Redundancy Optimizer (ZeRO) \cite{ZeroRedundancyOptimzer}. Here, 3D parallelism refers to Data Parallelism (DP) \cite{BenNun2019dp}, Pipeline Parallelism (PP) \cite{Gpipe, li2021terapipe} and Tensor Parallelism (TP) \cite{Megatron-LM}, which are often implemented to support extremely parallel execution of pre-training missions of LLMs across hundreds of GPUs. Also, tensor parallelism and pipeline parallelism are generally categorized under model parallelism. Data parallelism generally aims to partition one input data batch into mini-batches and distribute them to each GPU. This method enables parallel input data processing but requires a data-intensive All-reduce to aggregate the gradients at the end of backward propagation before updating each model replica. Pipeline parallelism further divides the model layers among workers and performs corresponding forward and backward computations on the micro-batches through a pipeline manner. This process involves mainly point-to-point activation passing from one device to another. Tensor parallelism focuses on splitting tensor computation among workers but also requires All-gathering and All-reducing tensors from all the devices. ZeRO reduces GPU memory footprint by partitioning model states among GPUs, further eliminates replications, and uses gather-based routine to reconstruct model states \cite{ZeroRedundancyOptimzer}.

\begin{figure*}[htbp]
    \centering
    \begin{subfigure}[b]{0.35\textwidth}
        \centering
        \includegraphics[width=\textwidth]{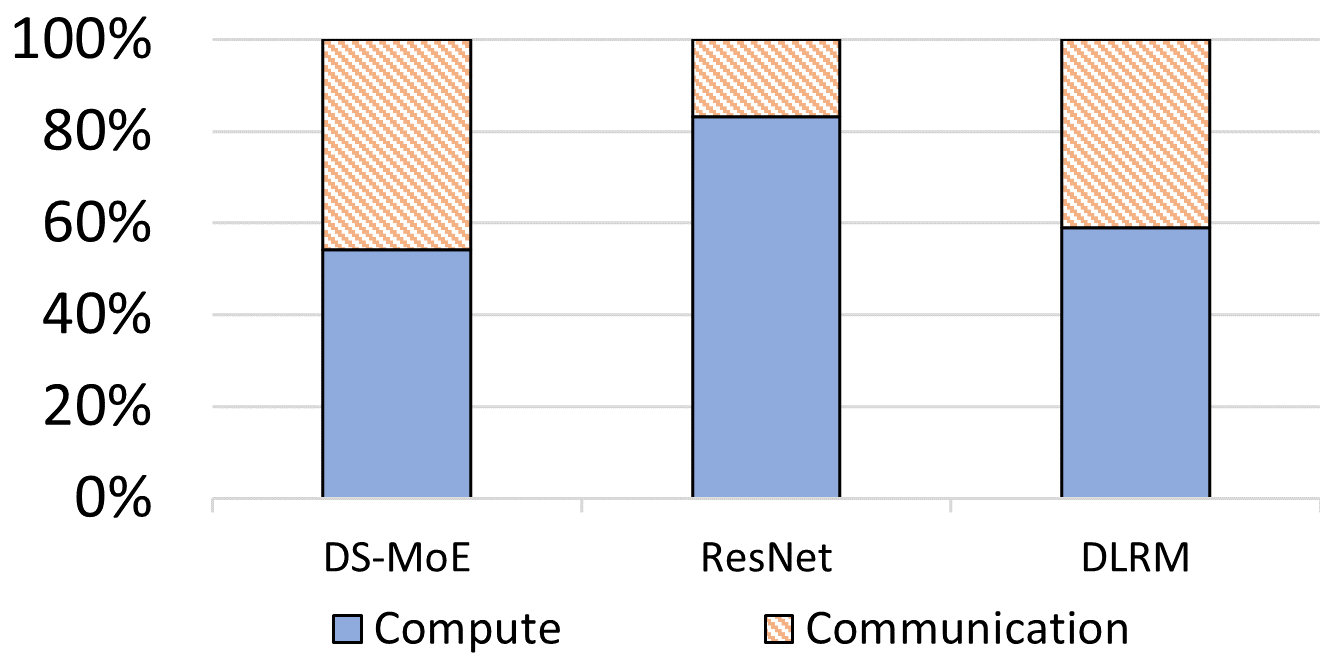}
        \caption{Proportion of computation to communication for distributed DL training}
        \label{fig:computevcomms}
    \end{subfigure}
    \hspace{4ex}
    \begin{subfigure}[b]{0.35\textwidth}
        \centering
        \includegraphics[width=\textwidth]{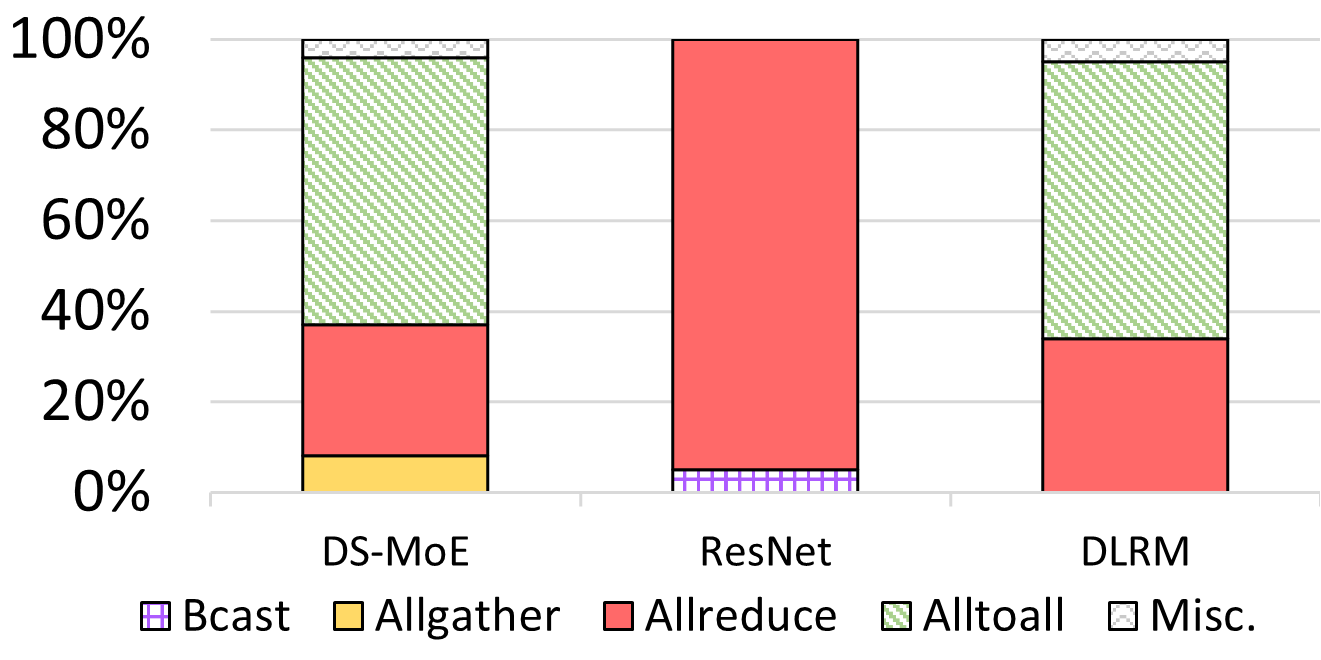}
        \caption{Breakdown of individual communication operations for distributed DL training}
        \label{fig:comms-breakdown}
    \end{subfigure}
    \caption{Communication profiling conducted in MCR-DL \cite{mcr-dl} for data-parallel and hybrid-parallel DL models.}
    \label{fig:comms-profile}
    \vspace{-2ex}
\end{figure*}

\subsection{Motivation}
\label{sec:motivation}

Applying 3D parallelism with ZeRO for DL model training imposes data-intensive collective communications within and across nodes. Given the limited interconnect bandwidth between GPU nodes, such large data transfer leads to drastic overhead. Profiling conducted by previous works \cite{mcr-dl} has also addressed that different parallelism strategy requires huge portions of communication and heterogeneous collective operations (see Figure \ref{fig:comms-profile}). Consequently, optimizing point-to-point and collective communication schemes became critical to mitigate such bottlenecks. Previous research has demonstrated that when co-designed with GPU-based compression libraries like ZFP \cite{ZFP} and MPC \cite{MPC}, GPU-aware MPI collectives were able to leverage interconnect bandwidth and stage outstanding throughput benefits \cite{Compression-IPDPS, compression-bcast, compression-redscat-allgather}. 

Conducting GPU-based compression techniques on data buffers drastically decreases the message size being communicated. However, representing high-precision data using lower precision always results in accuracy degradation, which is often observed within Deep Learning workloads. Prior researchers mainly experiment with small deep learning models to pick the most suitable compression rate for lossy ZFP library \cite{compression-redscat-allgather}. Yet, these models contain much fewer parameter numbers than LLMs, emitting smaller message sizes during transfer. \textbf{One key motivator of this work is to use compression-assisted MPI collectives to accelerate large language model training.}

\subsection{Challenges}
\label{sec:challenges}
This section addresses the following challenges:
\begin{enumerate}
    \item[\textbf{C1)}] The first challenge was determining the different model information communicated in parallelism stages. Pipeline parallelism, tensor parallelism, and data parallelism usually pose collective communications on various training parameters. Depending on the specific implementation, tensor parallelism \cite{Megatron-LM} usually calls All-reduce and All-gather on activations during forward passes and gradients during backward passes. Pipeline parallelism \cite{Gpipe,li2021terapipe} typically features point-to-point operations to pass activations and gradients from one stage to another. Finally, data parallelism requires All-reduce on the gradients of each data parallel rank and re-distributing the aggregated global gradients back to each model replica to perform subsequent updates \cite{anthony-segmentation, DP-Super-Resolution}. To overcome this challenge, we present a thorough illustration of the communication routines involved in 3D parallelism as well as in ZeRO stage 1 (Figure~\ref{fig:base-3D}~\ref{fig:zero1-base}). We further explain this demonstration in Section \ref{sec:design}

    \item[\textbf{C2)}] The second challenge is that compressing messages communicated during training depends on the parallelism stages and requires prudent design. We observed that na\"ively applying compression scheme to all parallelism stages introduced outstanding training speedup, but at the same time led to degradation in training quality (Section \ref{sec:naive-eval}). Current studies have shown that model parallelism has inherently different characteristics than data parallelism. In data parallelism, gradients communicated among different ranks are observed to be low-rank or sparse. Yet, in model parallelism where communicating activations is the bottleneck, activations are analyzed as dense \cite{bian2023does}. Therefore, applying the same compression intensity to both gradients and activations may lead to a loss of accuracy. Given these differences, compression intensity may vary to achieve a balanced solution that maximizes both throughput benefits and model training performance.

    \item[\textbf{C3)}] The third challenge we tackle in this paper originates from the fact that some model information, specifically gradients, is processed in both model-parallel and data-parallel stages. For example, both data parallelism and tensor parallelism feature the communication of gradients in their executions. When designing compression schemes, we should only apply aggressive compression to these gradients once, preventing over-extracting and hurting meaningful information, thus destroying accuracy. However, we cannot use lossless or high-precision methods towards these gradients since most negligible values will not be adequately extracted. Consequently, we need to avoid over-compression of gradients and maintain compression intensity. This challenge is being solved by applying different compression intensities towards different parallelism communication paths.
\end{enumerate}

\subsection{Solutions}
\label{sec:solution}
In this paper, we propose \textit{MZHybrid} and \textit{ZHybrid}, two hybrid compression schemes that utilize GPU-based compression on LLM training data, thus expediting the training process while maintaining acceptable model performance. We adopted MPI collectives co-designed with lossless MPC and lossy ZFP libraries to reduce the amount of data movement and leverage inter-node bandwidth. We analyzed different communication scenarios under 3D parallelism and ZeRO stage 1 optimization and designed an appropriate compression approach that considers the sparsity distinction between messages in the training process.

\subsection{Contributions}
\label{sec:contributions}
This paper contributes in the following manners:



\begin{enumerate}

    \item[\textbf{1)}] We experimented na\"ive compression schemes for both MPC and ZFP on modern large-scale HPC systems. We reported up to a 23.6\% increase in training samples per second and up to a 22.5\% increase in TFLOPS per GPU over non-compressed collective communications. We analyzed such schemes' benefits and shortcomings, leading to further design choices (Section \ref{sec:naive-eval}).

    \item[\textbf{2)}] We proposed and designed \textit{MZHybrid}, a hybrid compression scheme that utilizes lossless MPC for model-parallel communications and lossy ZFP rate for data-parallel communications. (Section \ref{sec:MZHybrid}) We also proposed and designed \textit{ZHybrid}, a hybrid compression scheme that utilizes different ZFP compression rates for communicating model-parallel and data-parallel messages. (Section \ref{sec:ZHybrid}).

    \item[\textbf{3)}] We evaluated \textit{MZHybrid} performance on modern large-scale HPC systems. We reported up to 4.4\% increase in training samples per second and up to 5.3\% increase in TFLOPS per GPU compared to non-compressed collective solutions. We demonstrated a significant improvement in model quality compared to the na\"ive ZFP scheme (Section \ref{sec:MZHybrid-eval}). We also evaluated \textit{ZHybrid} performance on modern large-scale HPC systems. We reported up to 17.3\% increase in training samples per second as well as up to 12.7\% increase in TFLOPS per GPU compared to non-compressed collective solutions. We also showed noticeable enhancement in model quality compared to na\"ive ZFP scheme (Section \ref{sec:ZHybrid-eval}).

    \item[\textbf{4)}] To the best of our knowledge, this is the first work that utilizes MPI collectives co-designed with GPU-based compression libraries to accelerate both model parallelism, data parallelism, and ZeRO communication (Section V).

\end{enumerate}

\vspace{-1.5ex}

\section{Background}
\label{sec:background}
\subsection{GPU-based Compression libraries and MPI}
The recent advancements in GPU technology, such as enhanced memory bandwidth, elevated core counts, and superior computation capabilities, have been a major factor in the adoption of GPU-based compression libraries. MPC \cite{MPC} is a lossless compression technique that leverages the identification of similarity between consecutive floating-point numbers to compress the data and achieve a high compression ratio. ZFP \cite{ZFP, ZFP-Analysis} is the state-of-the-art GPU-based compression library that supports high throughput read and write random access. These compression libraries have also been co-designed with MPI collective communication to achieve high-performance, on-the-fly message compression for modern, dense GPU clusters \cite{Compression-IPDPS}. Collective-level optimizations often avoid superfluous compression operations and utilization of GPU-kernels \cite{compression-bcast, compression-redscat-allgather, Compression-Alltoall}. 



\subsection{3D Parallelism and ZeRO}
This section covers backgrounds on 3D parallelism, which combines data-parallel, pipeline-parallel, and tensor-parallel. We will also discuss relevant works on ZeRO.

\subsubsection{Data Parallelism (DP)}

Data Parallelism \cite{BenNun2019dp} is a distributed DL technique that distributes training data across multiple GPUs with model replicas to perform parallel training steps. Data Parallelism can significantly improve throughput compared to single-node training, as evidenced by relevant applications \cite{anthony-segmentation, DP-Super-Resolution, scamp2023}. However, due to memory restrictions, DP has limitations with large, dense neural networks and high-resolution images.
\subsubsection{Model Parallelism (MP)}
Model parallelism overcomes the limitation of DP by splitting the model into layers and distributing it on different devices. Common approaches to achieve model parallelism include Pipeline Parallelism (PP) \cite{Gpipe, li2021terapipe} 
and Tensor Parallelism (TP) \cite{Megatron-LM}. PP uses inter-layer model parallelism with micro-batches executed in a pipeline, while TP employs sub-tensor splitting for parallel processing across multiple workers, optimizing tensor operations.

Higher-degree parallelism strategies emerged to further scale and harness the benefits of individual distributed models. MT-NLG \cite{smith2022using} uses 3D parallelism, a combination of DP, PP, and TP, to train a billion-parameter model, facilitating larger model sizes and improved training capabilities.

When analyzing the communication pattern for DP and MP, we observe that DP involves less frequent All-reduce operations on larger data sizes. At the same time, MP requires more frequent point-to-point operations on smaller data.

\subsubsection{ZeRO}

ZeRO (Zero Redundancy Optimizer) \cite{ZeroRedundancyOptimzer} provides a memory optimization technique and overcomes the memory limitation of DP and the scalability issue of MP. This is achieved by partitioning optimizer states, gradients, and model parameters across GPUs to eliminate redundant storage and GPU memory consumption. Efforts have also been made to offload certain training parameters to CPU and NVMe memory \cite{zero-infinity} and optimize communication overhead \cite{ZeRO++}. For example, ZeRO stage 1 maintains model weight and gradient copies on each GPU but partitions optimizer states among devices.

\subsection{Compression in Data Parallelism and Model Parallelism}
Data parallelism incurs communication overhead due to the All-reduce operation required for gradient aggregation. Various works have proposed gradient compression techniques for DP to address the overhead and improve training speed while maintaining accuracy \cite{signSGD, Deep-Grad-Compression}.
These works are based on the sense that most of the gradients communicated in DP are sparse, which means these data structures have most of their elements concentrated in a few dimensions. In contrast, the remaining dimensions contain negligible values. A comprehensive study \cite{bian2023does} has performed a low-rank analysis on the activation data in MP and concluded the opposite findings: activations in MP are dense, leading to its significance in preserving accuracy.

\section{Design}
\label{sec:design}

This section first details communication routines and messages in typical 3D parallelism and ZeRO stage 1 scenarios. Next, we will introduce the \textit{MZHybrid} scheme and the \textit{ZHybrid} scheme.

Figure \ref{fig:base-3D} illustrates a typical 3D parallelism setting split across eight global workers. This setting contains two DP ranks, two PP stages within each DP rank, and two TP degrees within each PP stage. We examine communication calls in each parallelism dimension in the following paragraphs. In Figure \ref{fig:base-3D}, the global batch is split into two mini-batches, and each goes into a DP rank. Each DP rank will produce its own local gradients after a forward and a backward pass on the model replicas. Then, the root rank will issue an All-reduce call on these local gradients and re-distribute the global aggregated gradients to each DP rank before updating the model replica parameters. The All-reduce call is graphed between DP ranks in the middle of Figure \ref{fig:base-3D}. Typically, this All-reduce is conducted upon sparse gradients. However, as model size increases, such collective communication gradually becomes a bottleneck, given the increase in message size. Next, we split the model replica into two PP stages within each DP rank. Each PP stage will include a subset of the network layers. For example, in Figure \ref{fig:base-3D}, we assume a model with eight layers, and we split them across two pipeline stages, with pipeline stage 0 having network layers 0-3 and pipeline stage 1 having network layers 4-7. Pipeline stages mainly call point-to-point communication routines like MPI\_Send \& MPI\_Recv to communicate gradients and activations with each other. These are also drawn between pipeline stages in Figure \ref{fig:base-3D}.

Next, we consider Tensor Parallelism. TP focuses on parallelizing matrix computations among workers. In this scenario, we parallelize computation workloads across two GPUs. For specific implementation, we refer to Megatron-LM \cite{Megatron-LM}. In their approach, they split GEMM operations in MLP blocks, Self-Attention blocks, and the output embedding layer among GPUs. This method requires two All-reduce primitives for a forward and a backward pass in a single transformer layer. For the output embedding layer, an All-reduce aggregates the different portions of the input embedding, and an All-gather will act to obtain GEMM outputs. Please refer to Figure \ref{fig:megatron-tp-allreduce} for details. In this parallelism dimension, the main communication primitives involved are All-reduce and All-gather—these primitives aggregate activations in forward passes and gradients in backward passes. We also illustrate these on the sides of pipeline stage blocks in Figure \ref{fig:base-3D}.

\begin{table}[htbp]
\centering
\resizebox{\columnwidth}{!}{%
\begin{tabular}{|c|c|c|c|c|}
\hline
\multicolumn{1}{|c|}{\textbf{MPI Collectives}} &
  \multicolumn{1}{c|}{\textbf{DP}} &
  \multicolumn{1}{c|}{\textbf{PP}} &
  \multicolumn{1}{c|}{\textbf{TP}} &
  \multicolumn{1}{c|}{\textbf{ZeRO stage 1}} \\ \hline
All-reduce     & \checkmark & $\times$                  & \checkmark                  & $\times$                  \\ \hline
All-gather     & $\times$                  & $\times$                  & \checkmark & \checkmark \\ \hline
Reduce-Scatter & $\times$                  & $\times$                  & $\times$                  & \checkmark \\ \hline
Point-to-point & $\times$                  & \checkmark & $\times$                  & $\times$                  \\ \hline
\end{tabular}%
}
\caption{MPI Collective communication in each parallelism stage.}
\label{tab:base-colls}
\end{table}

\begin{figure*}[htbp]
\vspace{-2ex}
\centering
    \includegraphics[width=0.6\textwidth]{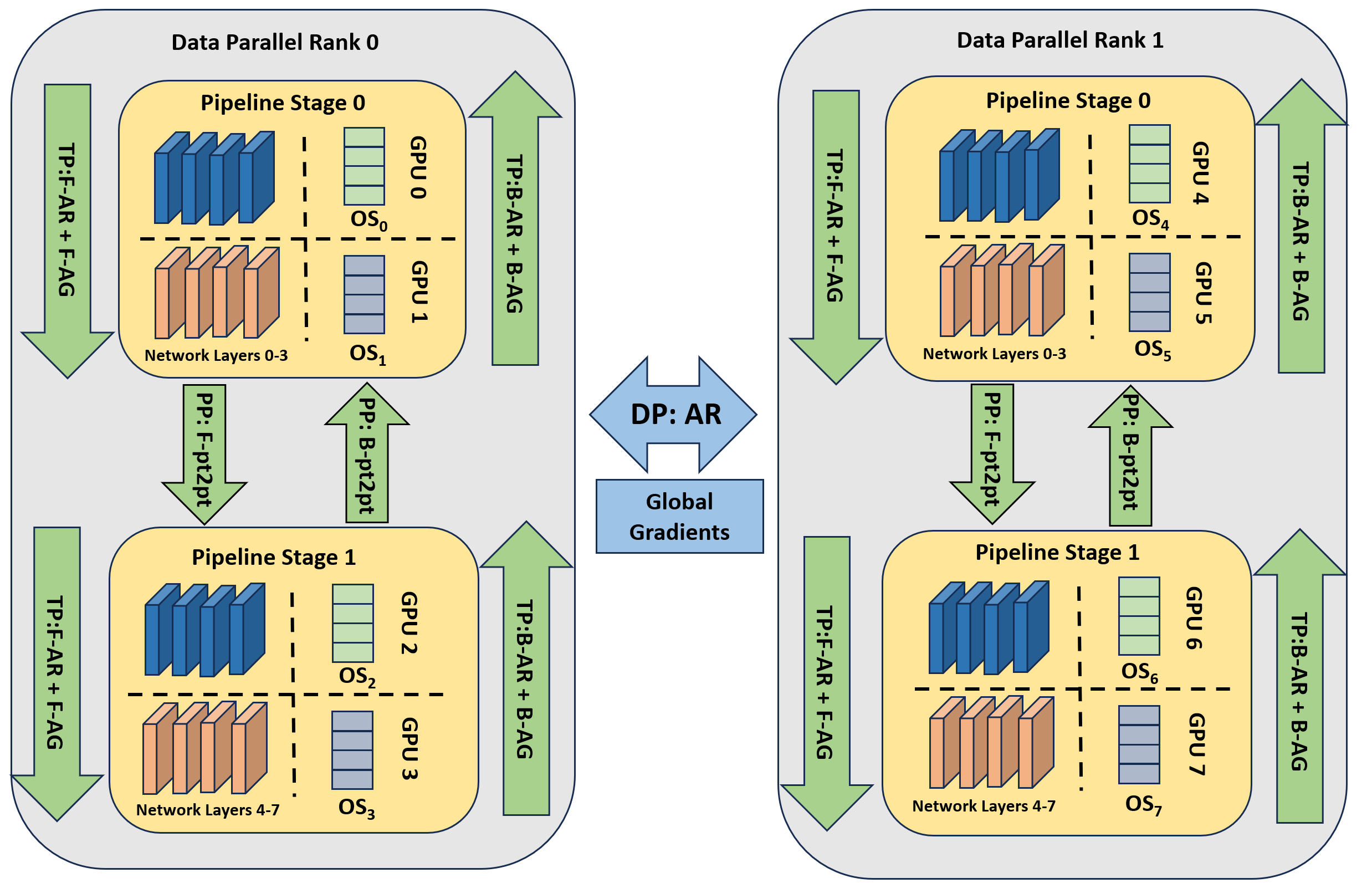}
    \caption{A base 3D parallelism setting with a total of 8 GPUs. This setting contains two data parallel ranks, two pipeline stages and two tensor parallel degree. Collective operations for each parallelism degree are detailed. Optimizer states are also split across 8 workers. \textit{DP}: Data Parallelism, \textit{PP}: Pipeline Parallelism, \textit{TP}: Tensor Parallelism, \textit{OS}: optimizer state shards with device index as subscript. \textit{AR}: All-reduce operations. \textit{AG}: All-gather operations. F and B prefix indicates forward passes and backward passes respectively.} 
    \label{fig:base-3D}
    \vspace{-2ex}
\end{figure*}

\begin{figure}[htbp]
\vspace{-2ex}
\centering
    \includegraphics[width=\columnwidth]{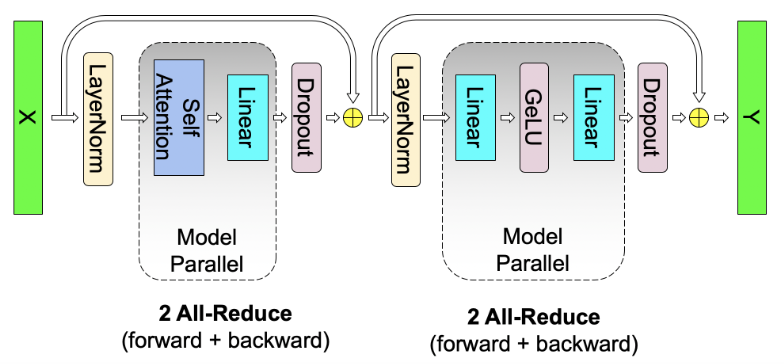}
    \caption{Megatron-LM: communication primitives involved in a single transformer layer \cite{Megatron-LM}.} 
    \label{fig:megatron-tp-allreduce}
    \vspace{-2ex}
\end{figure}

\begin{figure}[htbp]
\vspace{-5ex}
\centering
    \includegraphics[width=0.8\linewidth]{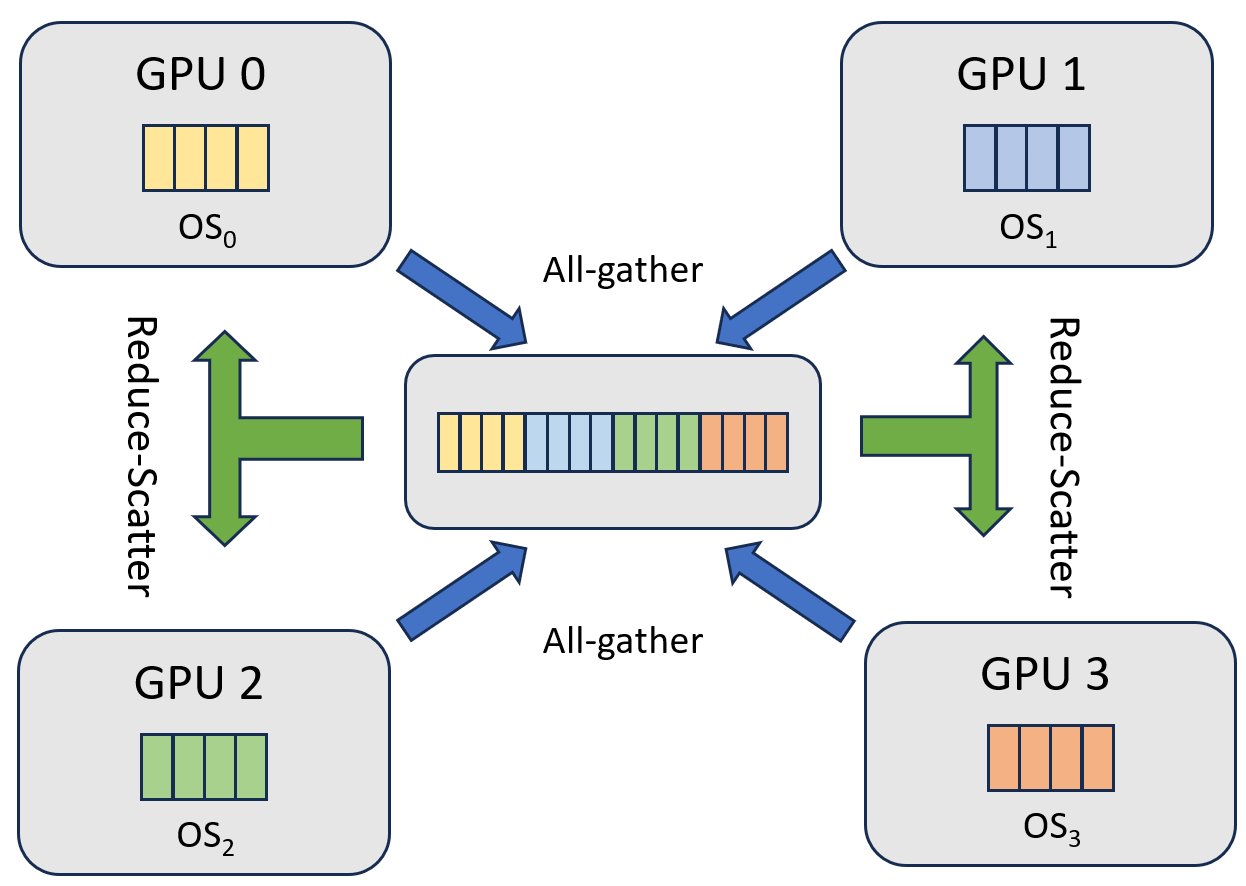}
    \caption{ZeRO: communication primitives involved in ZeRO stage 1 over 4 DP ranks(GPUs). Each DP rank will receive their optimizer state shard and update the according portion of model weights. In our experiments, ZeRO stage 1 is integrated into 3D parallelism, here we use a separate graph to demonstrate communication routines for clarity. OS: optimizer state partitions with GPU number as index.} 
    \label{fig:zero1-base}
    \vspace{-2ex}
\end{figure}

Next, we detail about ZeRO stage 1 communications. ZeRO stage 1 optimizes GPU memory consumption by partitioning optimizer states over several workers. Optimizers like Adam require considerable per-parameter information and can acquire twice as much memory as a complete model. ZeRO stage 1 only keeps a fraction of optimizer states on each GPU. Each worker will only perform weight updates on the portion related to its optimizer state shards. After each the model parameters are updated, an All-gather will be called to collect the updated parameters from all the DP ranks. We exemplify a ZeRO stage 1 scenario with optimizer states split across 4 GPUs in Figure \ref{fig:zero1-base}. Finally, in Table \ref{tab:base-colls}, we listed all the MPI collective communication per parallelism stage.

\subsection{\textit{MZHybrid}: MPC for MP + ZFP for DP}
\label{sec:MZHybrid}


\begin{table}[htbp]
\centering
\resizebox{\columnwidth}{!}{%
\begin{tabular}{|c|c|c|}
\hline
\multicolumn{1}{|c|}{\textbf{MZHybrid}} & \multicolumn{1}{c|}{\textbf{MPI Collectives}} & \multicolumn{1}{c|}{\textbf{Compression Schemes}} \\ \hline
DP                            & All-reduce     & ZFP \\ \hline
PP                            & Point-to-point & MPC \\ \hline
\multirow{2}{*}{TP}           & All-reduce     & MPC \\ \cline{2-3} 
                              & All-gather     & MPC \\ \hline
\multirow{2}{*}{ZeRO stage 1} & All-gather     & MPC \\ \cline{2-3} 
                              & Reduce-Scatter & MPC \\ \hline
\end{tabular}%
}
\caption{MZHybrid: Compression scheme specified for each collective communication}
\label{tab:mzhybrid-colls}
\end{table}

In this section, we introduce the first hybrid compression scheme: \textit{MZHybrid}. \textit{MZHybrid} uses lossless MPC scheme for MP communication and lossy ZFP scheme for DP communication. 

\begin{figure*}[htbp]
\vspace{-2ex}
\centering
    \includegraphics[width=0.75\textwidth]{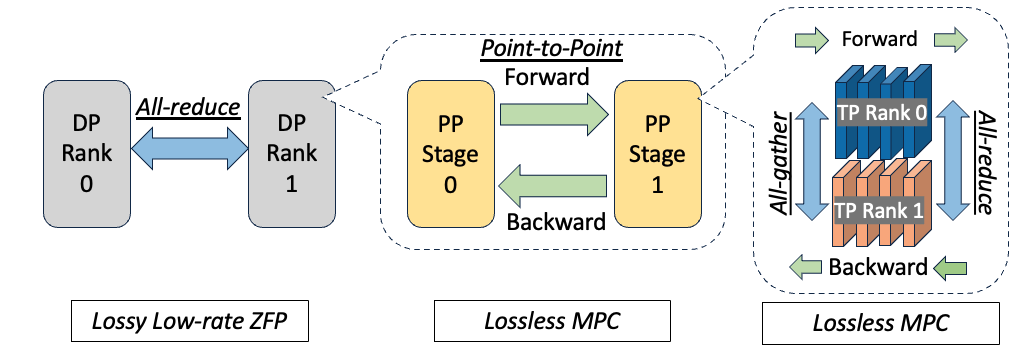}
    \caption{\textit{MZHybrid}: 3D Parallelism communication settings. Collective operations as well as compression schemes for each parallelism degree are detailed. We force lossless MPC for TP and PP while lossy ZFP for DP.} 
    \label{fig:mzhybrid-3d}
    \vspace{-2ex}
\end{figure*}


We provide illustrations using \textit{MZHybrid} under the typical 3D parallelism scenario in Figure \ref{fig:mzhybrid-3d}. We enforce lossless MPC for All-reduces in TP, point-to-point sends\&recvs in PP, and lossy ZFP for All-reduce between DP ranks. For ZeRO stage 1 under \textit{MZHybrid}, we enforce lossless MPC for communications. Given that activations communicated in MP are mostly dense (contrary to gradients) \cite{bian2023does}, we want to preserve precision on these data to maintain model training performance. For communicating large and mostly sparse gradients between DP ranks, we apply an aggressive lossy ZFP scheme. It is worth mentioning that in MP settings, communication on gradients also exists during backward passes. To avoid over-compressing gradients, we apply MPC schemes to those gradients. We use Table \ref{tab:mzhybrid-colls} to specify our compression scheme choice for each collective involved for \textit{MZHybrid}. We also experimented with different rates of ZFP under \textit{MZHybrid}, please refer to \ref{sec:MZHybrid-eval}.

\subsection{\textit{ZHybrid}: high-rate-ZFP for MP + low-rate-ZFP for DP}
\label{sec:ZHybrid}


This section presents the second hybrid compression scheme: \textit{ZHybrid}. \textit{ZHybrid} adopts a lossy ZFP scheme for all communication, including 3D parallelism stages and ZeRO stage 1. However, for different parallelism ranks, we apply different rates of ZFP. Since high-rate ZFP are better at preserving accuracy \cite{ZFP-Analysis}, we apply them towards MP units and ZeRO stage 1. For DP, We apply low-rate ZFP to large and sparse gradient reduction to eliminate negligible values. For experiments with different rates of ZFP under \textit{ZHybrid}, please refer to \ref{sec:ZHybrid-eval}. We use Table \ref{tab:zhybrid-colls} to specify our compression scheme choice for each collective involved for \textit{ZHybrid}.

\begin{table}[htbp]
\vspace{-3ex}
\centering
\resizebox{\columnwidth}{!}{%
\begin{tabular}{|c|c|c|}
\hline
\multicolumn{1}{|c|}{\textbf{ZHybrid}} & \multicolumn{1}{c|}{\textbf{MPI Collectives}} & \multicolumn{1}{c|}{\textbf{Compression Schemes}} \\ \hline
DP                            & All-reduce     & low-rate ZFP  \\ \hline
PP                            & Point-to-point & high-rate ZFP \\ \hline
\multirow{2}{*}{TP}           & All-reduce     & high-rate ZFP \\ \cline{2-3} 
                              & All-gather     & high-rate ZFP \\ \hline
\multirow{2}{*}{ZeRO stage 1} & All-gather     & high-rate ZFP \\ \cline{2-3} 
                              & Reduce-Scatter & high-rate ZFP \\ \hline
\end{tabular}%
}
\caption{ZHybrid: Compression scheme specified for each collective communication}
\label{tab:zhybrid-colls}
\end{table}

\begin{figure*}[htbp]
\vspace{-1ex}
\centering
    \includegraphics[width=0.75\textwidth]{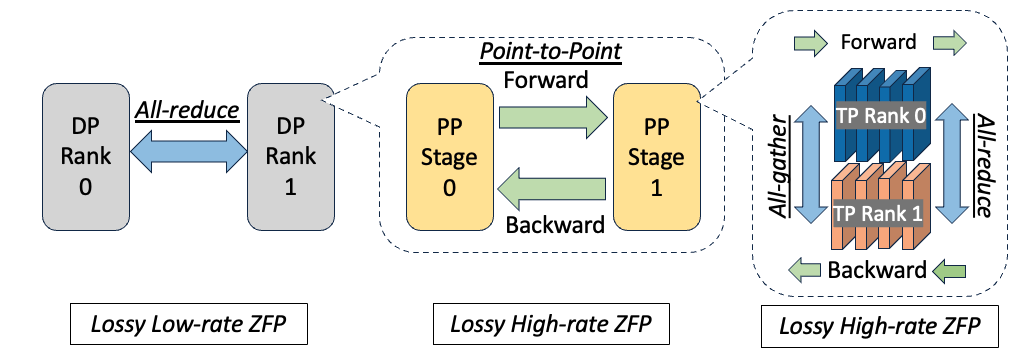}
    \caption{\textit{ZHybrid}: 3D Parallelism communication settings. Collective operations as well as compression schemes for each parallelism degree are detailed. We force high-rate ZFP for TP and PP while low-rate ZFP for DP.} 
    \label{fig:zhybrid-3d}
    \vspace{-2ex}
\end{figure*}




\section{Evaluation}
\label{sec:evalutaion}



In this section, we evaluate different hybrid compression schemes for LLM training in terms of training throughput (TFLOPS) and loss. We conducted experiments on the Lassen supercomputer hosted at Lawrence Livermore National Laboratory \cite{lassen}. The cluster comprises 792 GPU nodes, each with four 16 GB memory NVIDIA Tesla V100 GPUs and two 44-core IBM Power 9 CPUs. Inter-node connection is established through Mellanox Infiniband EDR with a bandwidth of 100Gb/s, and the 4 GPUs on each node are split into two pairs connected via NVLINK (Table \ref{tab:lassen-config}). Experiments on more advanced GPU architecture (A100, H100) would be beneficial, yet we believe that the core findings and narrative would largely remain the same. 

\begin{table}
\centering
\resizebox{\columnwidth}{!}{%
\begin{tabular}{|c|c|}
\hline
\textbf{CPU}          & IBM Power9 44 Cores/Node   \\ \hline
\textbf{Memory}       & 256 GB                     \\ \hline
\textbf{GPU} & \begin{tabular}[c]{@{}c@{}}4 NVIDIA Tesla V100 (64 GB Memory)\\ with NVLINK\end{tabular} \\ \hline
\textbf{Interconnect} & Mellanox IB EDR (100 Gb/s) \\ \hline
\end{tabular}%
}
\caption{Lassen cluster configuration}
\label{tab:lassen-config}
\end{table}

\subsection{Software Libraries}
\label{sec:software-env}
We invoked compression-assisted reduce-scatter-allgather All-reduce based on compression-assisted reduce-scatter and all-gather implemented on top of MVAPICH2-GDR 2.3.7 \cite{mvapich} for all training experiments. We chose the GPT-NeoX library given its open-sourced documentation and implementation of LLM parallel training procedures. This library was implemented based on Megatron-LM \cite{Megatron-LM} and DeepSpeed \cite{Rasley2020DeepSpeedSO} to support 3D parallelism. Furthermore, it has also been augmented with various novel optimizations, including ZeRO \cite{ZeroRedundancyOptimzer}, etc. We compiled PyTorch v1.13.1 and the latest DeepSpeed from source with GPU-aware MPI support.

\subsection{Training Configuration}
\label{sec:train-config}

 The inter-node interconnect is InfiniBand-EDR-100Gb/s, and the 4 GPUs on each node are split into two pairs connected via NVLINK. We agree that experiments on more advanced GPU architecture would be beneficial, but the fundamental storyline would be unchanged.
 
We selected the largest language model checkpointed in the GPT-NeoX library - GPT-NeoX-20B. Due to a lack of resources to train the original foundation model, it was necessary to conduct fine-tuning using a more constrained dataset. Specifically, the model was fine-tuned on 'Books3', a subset of the 'Pile' dataset developed by EleutherAI \cite{gao2020pile}. We set up the model using the same hyperparameters in the original paper \cite{black2022gptneox20b}. We trained the model for 4000 steps. We changed the parallelism settings to match the Lassen GPU node configuration. We set the pipeline parallelism degree to be 6 across nodes and model parallelism degree to be 4 within nodes. This makes up a base training environment among 24 GPUs for one model replica. We use a gradient accumulation step of 1, a training micro batch size per GPU of 4, and scaled training up to 192 V100 GPUs. For optimizer settings, we adopted the Adam optimizer with beta values of 0.9 and 0.95 as well as an epsilon of 1.0e-8. We also enabled ZeRO optimizer to distribute optimizer states across devices to reduce memory consumption. 

\subsection{Na\"ive ZFP Compression Scheme}
\label{sec:naive-eval}
We first forced na\"ive lossy ZFP compression schemes to all parallelism. We conclude our results of testing na\"ive ZFP(rate:8) and ZFP (rate:16) by reporting training throughput in two aspects: samples per second and TFLOPS per V100 GPU. We also documented test loss on Books3 for model performance validation. In Figure \ref{fig:naive-samples-sec}, when compared to default MVAPICH2-GDR implementations, conducting ZFP compression techniques showcased a \textbf{23.6\%} increase in samples per second for rate:8 and a \textbf{15.4\%} respectively for rate:16 on 192 V100 GPUs. In Figure \ref{fig:naive-tflops}, we observed a \textbf{22.5\%} increase in TFLOPS per V100 GPU for rate:8 and a \textbf{11.14\%} increase in that for rate:16 on 192 V100 GPUs. The throughput benefit difference between ZFP rate:8 and rate:16 is expected since, with a lower rate, we are extracting out more information, thus leading to more bandwidth being freed and better throughput. We continued to evaluate test loss on the trained model; we discern some degradation in the loss curves in Figure \ref{fig:naive-loss}. Compared to baseline showing a steep decrease in test loss, na\"ively compressing all messages using ZFP produces flatter loss curves and, eventually, larger loss values. It is also worth noting that a larger ZFP compression rate yields loss curves and values closer to baseline since less model information is being compressed during message passing.  

\begin{figure*}[hbtp]
    \centering
    \begin{subfigure}[b]{0.32\textwidth}
        \centering
        \includegraphics[width=\textwidth]{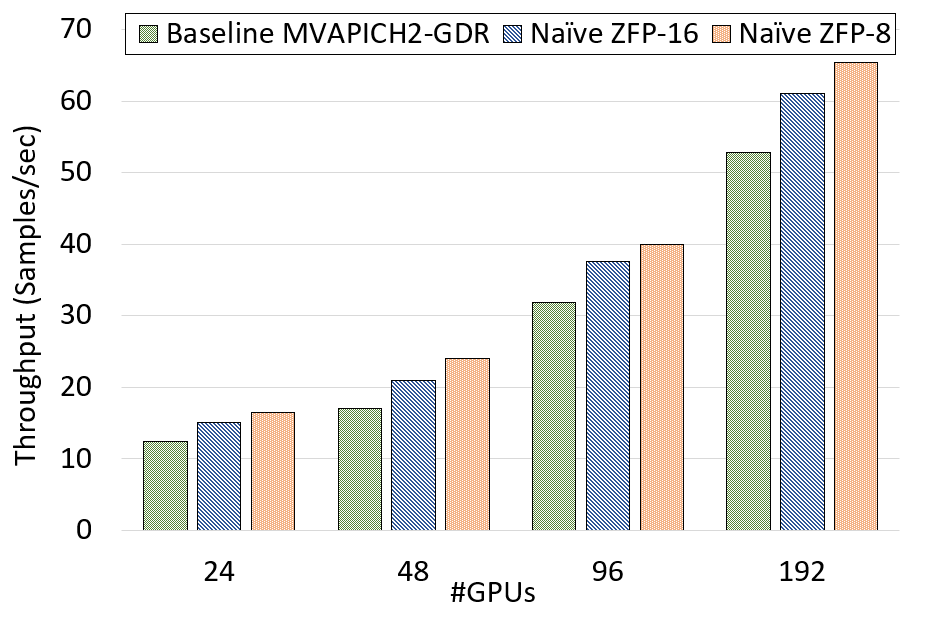}
        \caption{Naive ZFP: Training samples per second}
        \label{fig:naive-samples-sec}
    \end{subfigure}
    \begin{subfigure}[b]{0.32\textwidth}
        \centering
        \includegraphics[width=\textwidth]{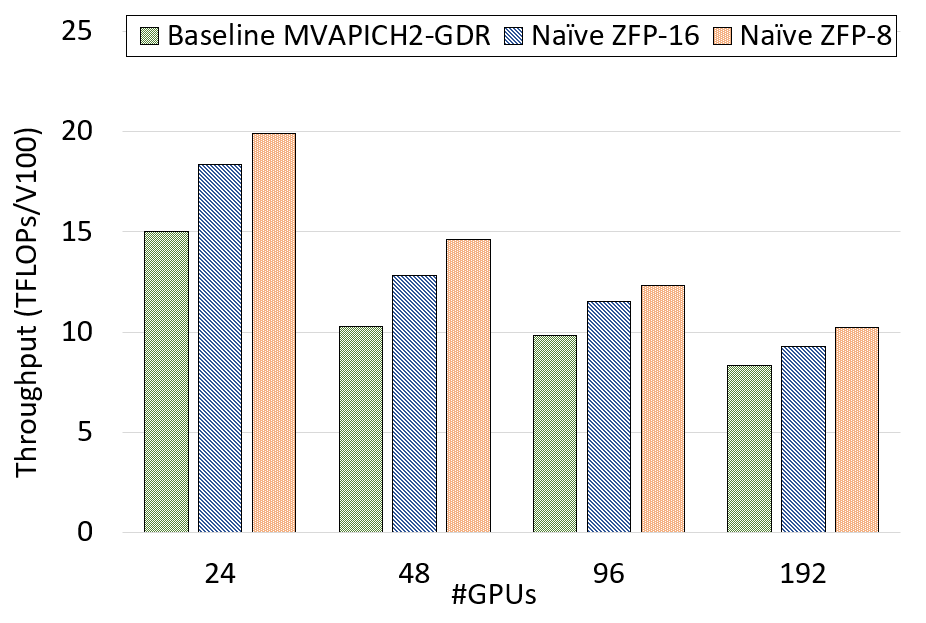}
        \caption{Na\"ive ZFP: TFLOPS per GPU}
        \label{fig:naive-tflops}
    \end{subfigure}
    \begin{subfigure}[b]{0.32\textwidth}
        \centering
        \includegraphics[width=\textwidth]{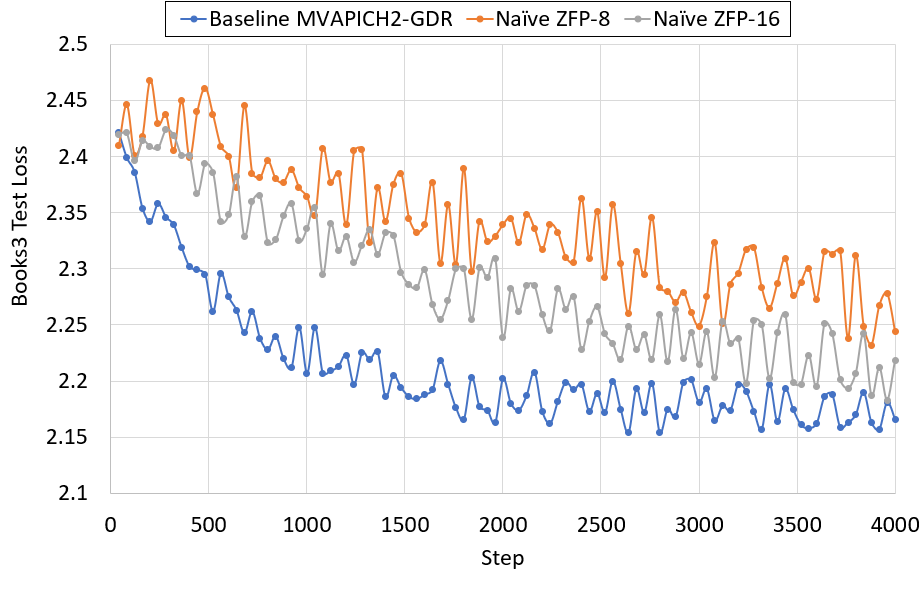}
        \caption{Na\"ive ZFP: Books3 test loss}
        \label{fig:naive-loss}
    \end{subfigure}
    \caption{Na\"ive ZFP Compression Scheme}
    \label{fig: naive-zfp}
\end{figure*}

\subsection{Na\"ive MPC Compression Scheme}
\label{sec:mpc-eval}
Secondly, we switched to using pure lossless MPC compression. We similarly applied MPC to all parallelism in Figure \ref{fig:naive-mpc-samples-sec} and Figure \ref{fig:naive-mpc-tflops}, and we report that using MPC didn't show significant throughput benefits. Nevertheless, in Figure \ref{fig:naive-mpc-loss}, we observed that when comparing to baseline loss curves, MPC elicited a better ability to match desired model performance and hardly produced any degradation. This observation is anticipated since MPC compression is a lossless scheme and proficient in preserving model data precision. 

\begin{figure*}[hbtp]
    \centering
    \begin{subfigure}[b]{0.32\textwidth}
        \centering
        \includegraphics[width=\textwidth]{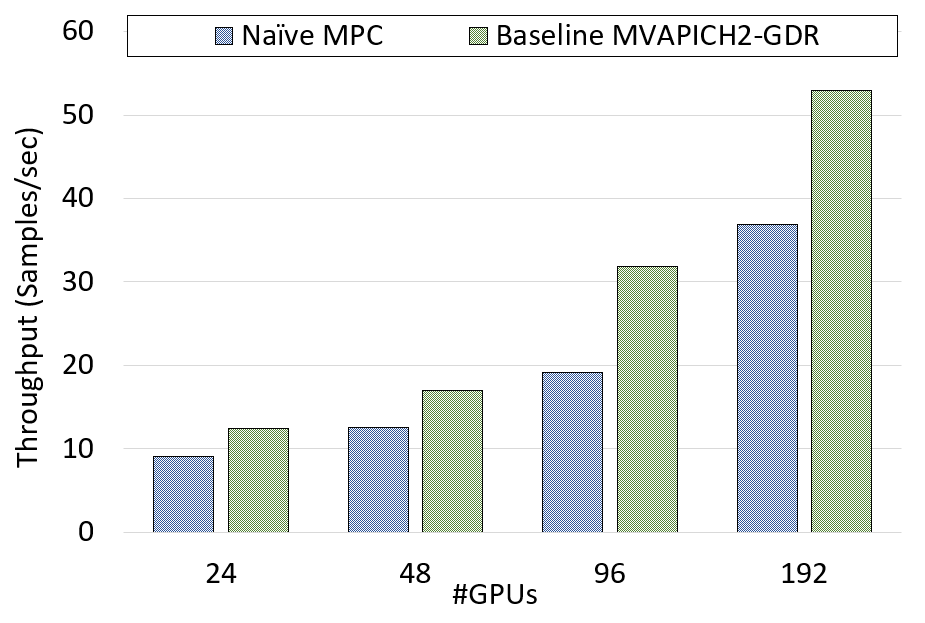}
        \caption{Na\"ive MPC: Training samples per second}
        \label{fig:naive-mpc-samples-sec}
    \end{subfigure}
    \begin{subfigure}[b]{0.32\textwidth}
        \centering
        \includegraphics[width=\linewidth]{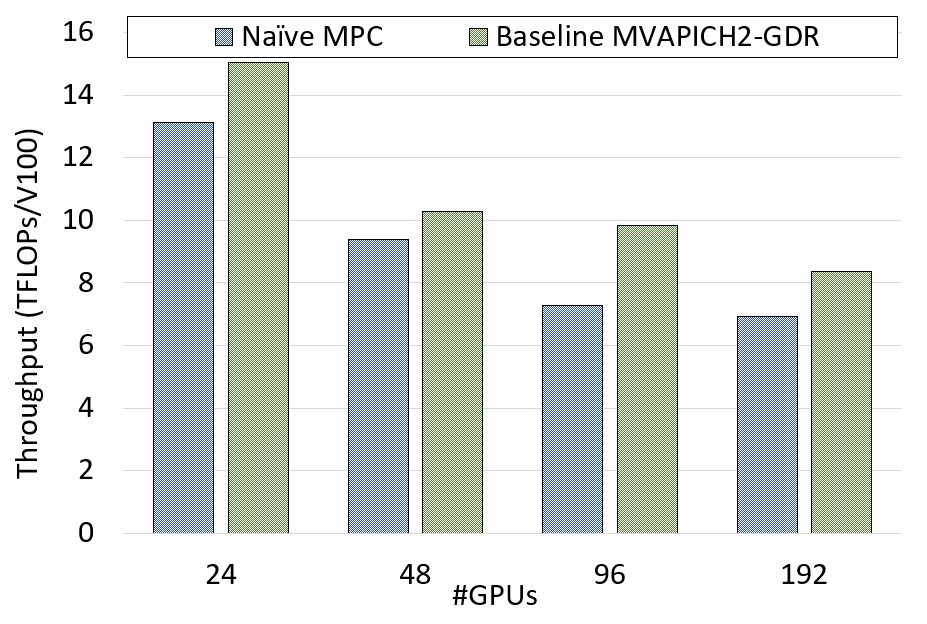}
        \caption{Na\"ive MPC: TFLOPS per GPU} 
        \label{fig:naive-mpc-tflops}
    \end{subfigure}
    \begin{subfigure}[b]{0.32\textwidth}
        \centering
        \includegraphics[width=\linewidth]{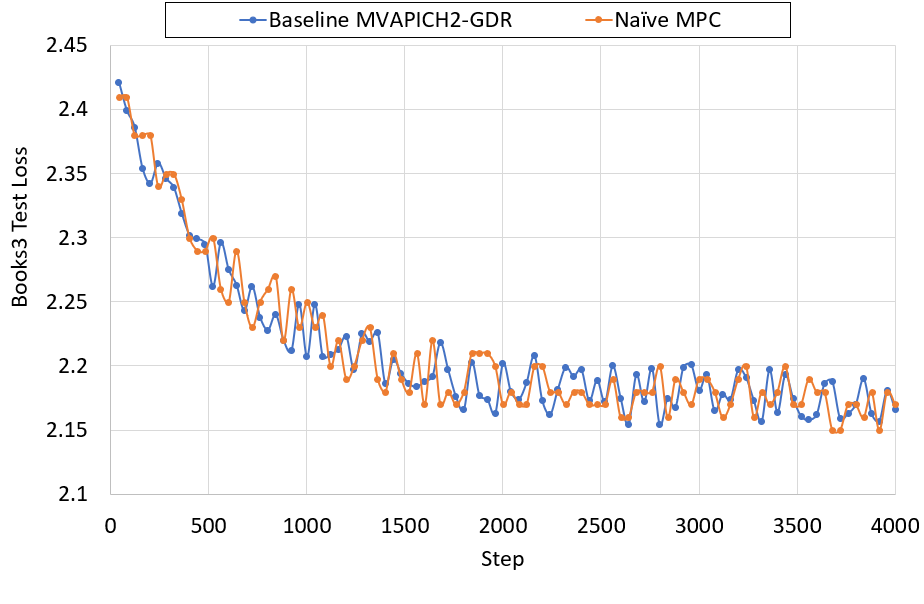}
        \caption{Na\"ive MPC: Books3 test loss} 
        \label{fig:naive-mpc-loss}
    \end{subfigure}
    \caption{Na\"ive MPC Compression Scheme}
    \label{fig: naive-mpc}
\end{figure*}

\subsection{\textit{MZHybrid}}
\label{sec:MZHybrid-eval}
In this section, we evaluate the \textit{MZHybrid} compression scheme—applying lossless MPC in MP communication and lossy ZFP in DP communication. We recorded results for ZFP rate:8 and ZFP rate:16. In Figure \ref{fig:mzhybrid-samples-sec}, we saw that using ZFP rate:8 together with MPC, training samples per sec showed a \textbf{4.4\%} increase on 192 GPUs when compared to MVAPICH2-GDR baseline. For TFLOPS per GPU, we demonstrated in Figure \ref{fig:mzhybrid-tflops} a \textbf{5.3\%} raise when training across 192 GPUs. When it comes to model performance, we plotted loss curves with MZHybrid against na\"ive ZFP approach. Figure \ref{fig:mzhybrid-loss} staged that \textit{MZHybrid} significantly reduces loss values when compared to na\"ive schemes for both ZFP rate:8 and ZFP rate:16 (Figure \ref{fig:naive-loss}). We anticipate such observations, although MPC struggles at providing throughput benefits for large message size, the speed-up provided by ZFP offsets such shortcoming. At the same time, the loss curve showed that incorporating MPC significantly improves model performance.
\begin{figure*}[hbtp]
    \centering
    \begin{subfigure}[b]{0.32\textwidth}
        \centering
        \includegraphics[width=\linewidth]{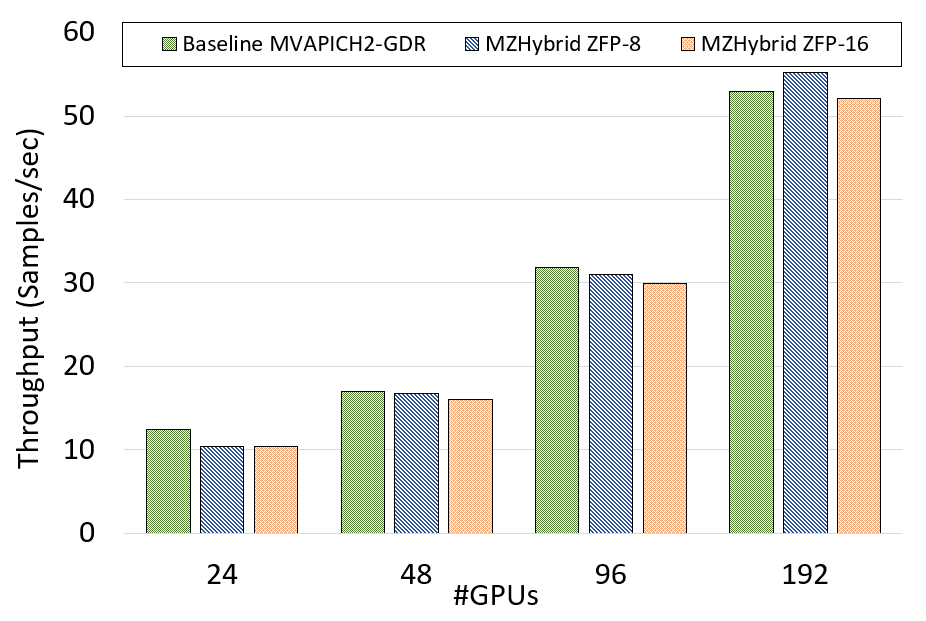}
        \caption{MZHybrid: Training samples per second} 
        \label{fig:mzhybrid-samples-sec}
    \end{subfigure}
    \begin{subfigure}[b]{0.32\textwidth}
        \centering
        \includegraphics[width=\linewidth]{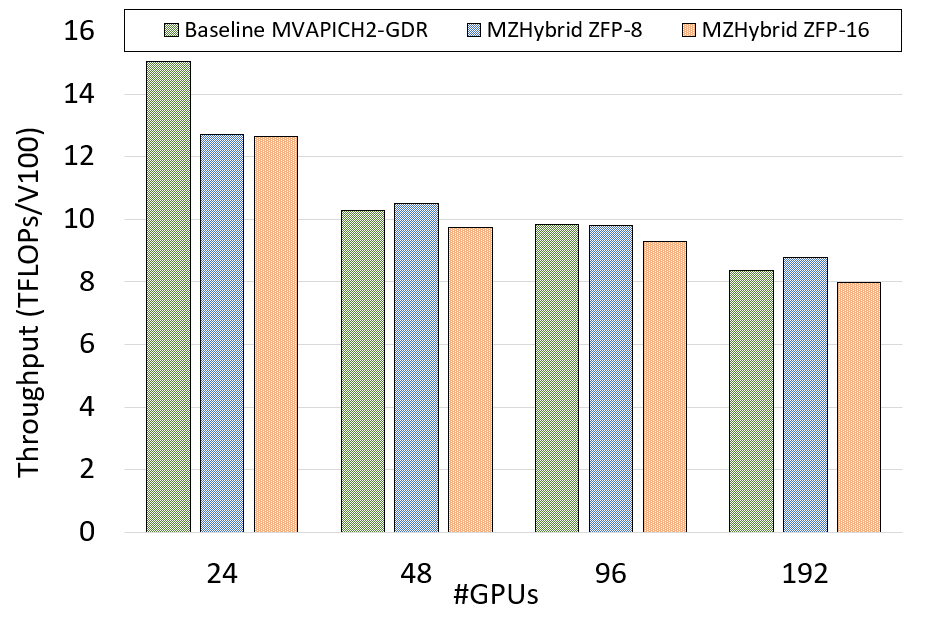}
        \caption{MZHybrid: TFLOPS per GPU} 
        \label{fig:mzhybrid-tflops}
    \end{subfigure}
    \begin{subfigure}[b]{0.32\textwidth}
        \centering
         \includegraphics[width=\linewidth]{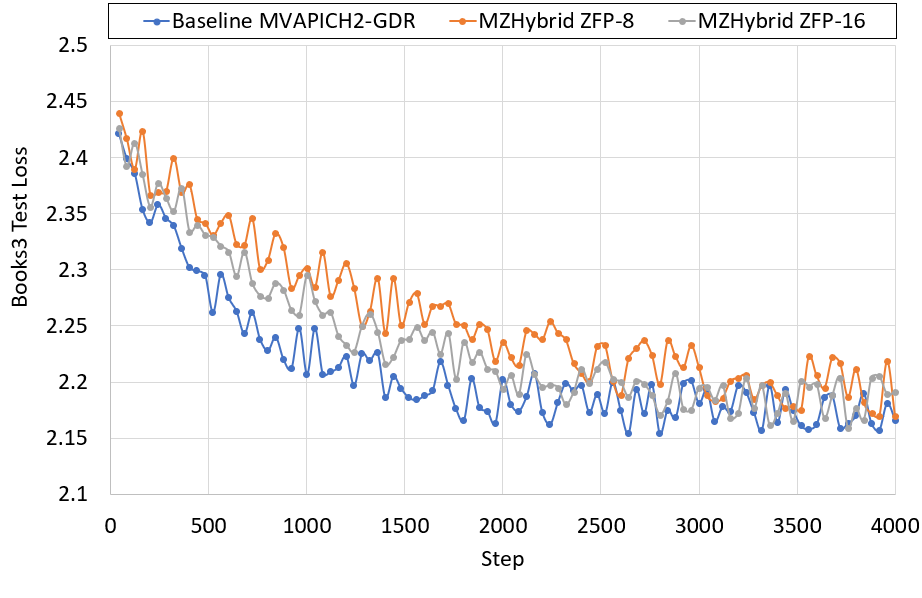}
        \caption{MZHybrid: Books3 test loss} 
        \label{fig:mzhybrid-loss}
    \end{subfigure}
    \caption{MZHybrid Compression Scheme}
    \label{fig: MZHybrid}
\end{figure*}

\subsection{\textit{ZHybrid}}
\label{sec:ZHybrid-eval}

We continued experimenting with \textit{ZHybrid}: using different ZFP compression rates for MP and DP stages. We conducted experiments on two cases: one is ZFP(rate:24) for MP and ZFP(rate:8) for DP, and the second one is ZFP(rate:16) for MP and ZFP(rate:8) for DP. While applying ZFP(rate:16) for MP, we observed a \textbf{20.4\%} increase in training samples per second and a \textbf{20.6\%} increase in TFLOPS per GPU. When considering ZFP(rate:24) for MP, we also see a \textbf{17.3\%} increase in training samples per second and a \textbf{12.7\%} increase in TFLOPS per GPU (Figure \ref{fig:zhybrid-samples-sec}, \ref{fig:zhybrid-tflops}). Then we compared \textit{ZHybrid} against na\"ive ZFP (Figure 11) in terms of test loss and observed lower final loss values, which translates to better model quality. Increasing the ZFP rate for MP communication improves model performance as expected. 

\begin{figure*}[hbtp]
    \centering
    \begin{subfigure}[b]{0.32\textwidth}
        \centering
        \includegraphics[width=\linewidth]{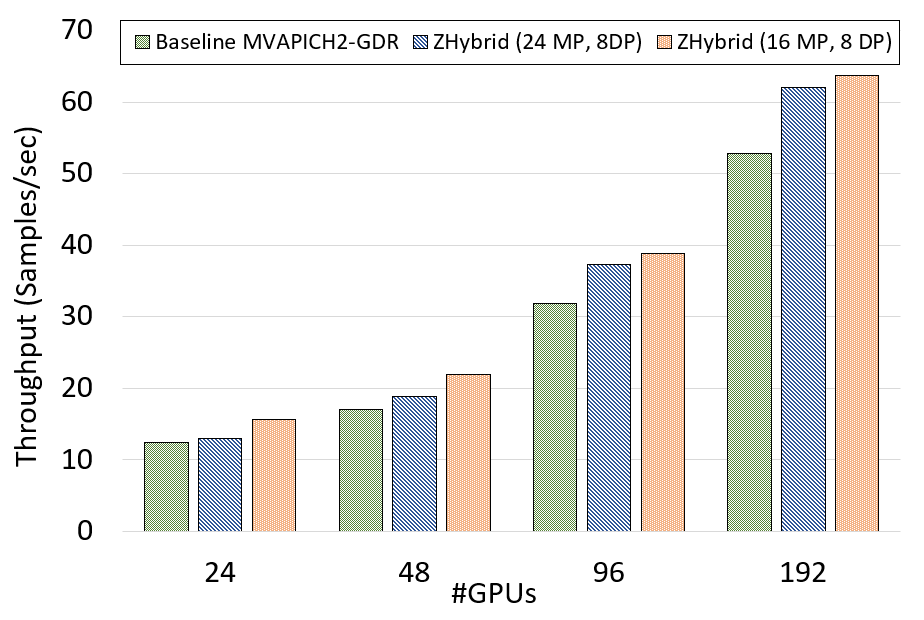}
        \caption{ZHybrid: Training samples per second} 
        \label{fig:zhybrid-samples-sec}
    \end{subfigure}
    \begin{subfigure}[b]{0.32\textwidth}
        \centering
        \includegraphics[width=\linewidth]{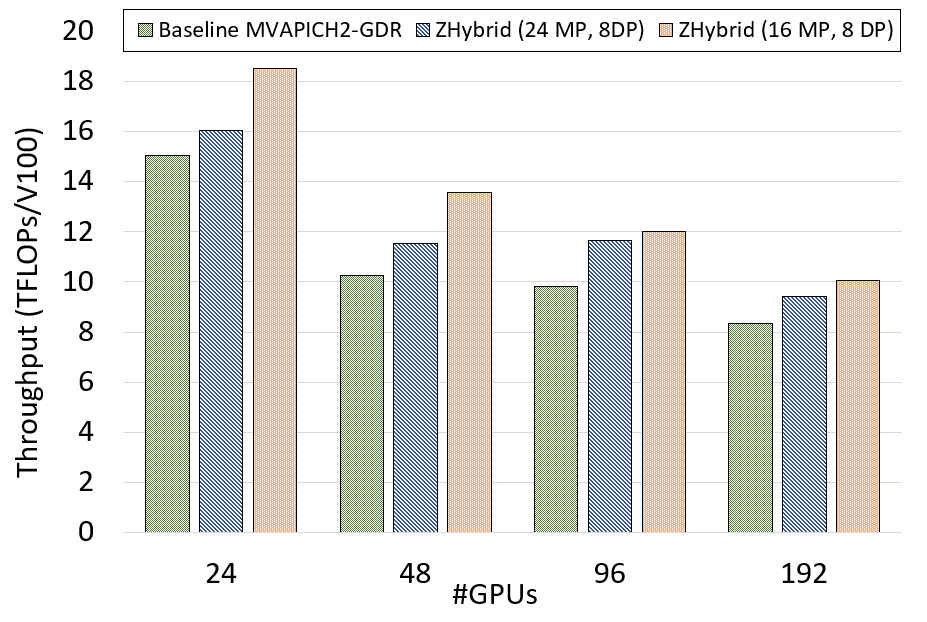}
        \caption{ZHybrid: TFLOPS per GPU} 
        \label{fig:zhybrid-tflops}
    \end{subfigure}
    \begin{subfigure}[b]{0.32\textwidth}
        \centering
         \includegraphics[width=\linewidth]{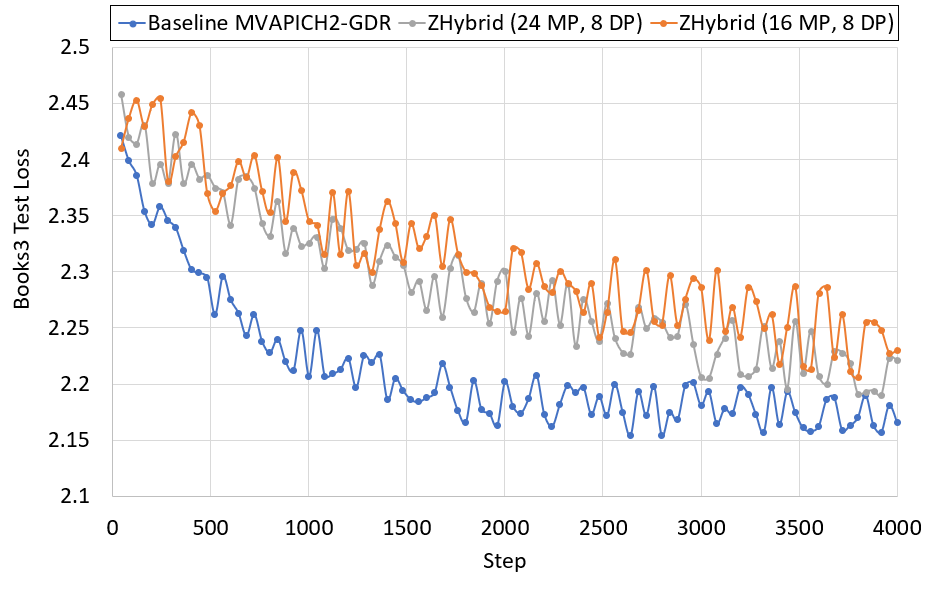}
        \caption{ZHybrid: Books3 test loss} 
        \label{fig:zhybrid-loss}
    \end{subfigure}
    \caption{ZHybrid Compression Scheme}
    \label{fig:ZHybrid}
\end{figure*}

Both evaluated \textit{ZHybrid} cases(rate:24 MP \& rate:16 MP) staged lower loss values over naïve ZFP solution. Also, the moving average of the loss is higher for \textit{ZHybrid}(rate:16 MP), which conforms with our expectation that lower ZFP rates lead to higher overall loss landscape.

\begin{figure}[htbp]
\centering
    \includegraphics[width=\columnwidth]{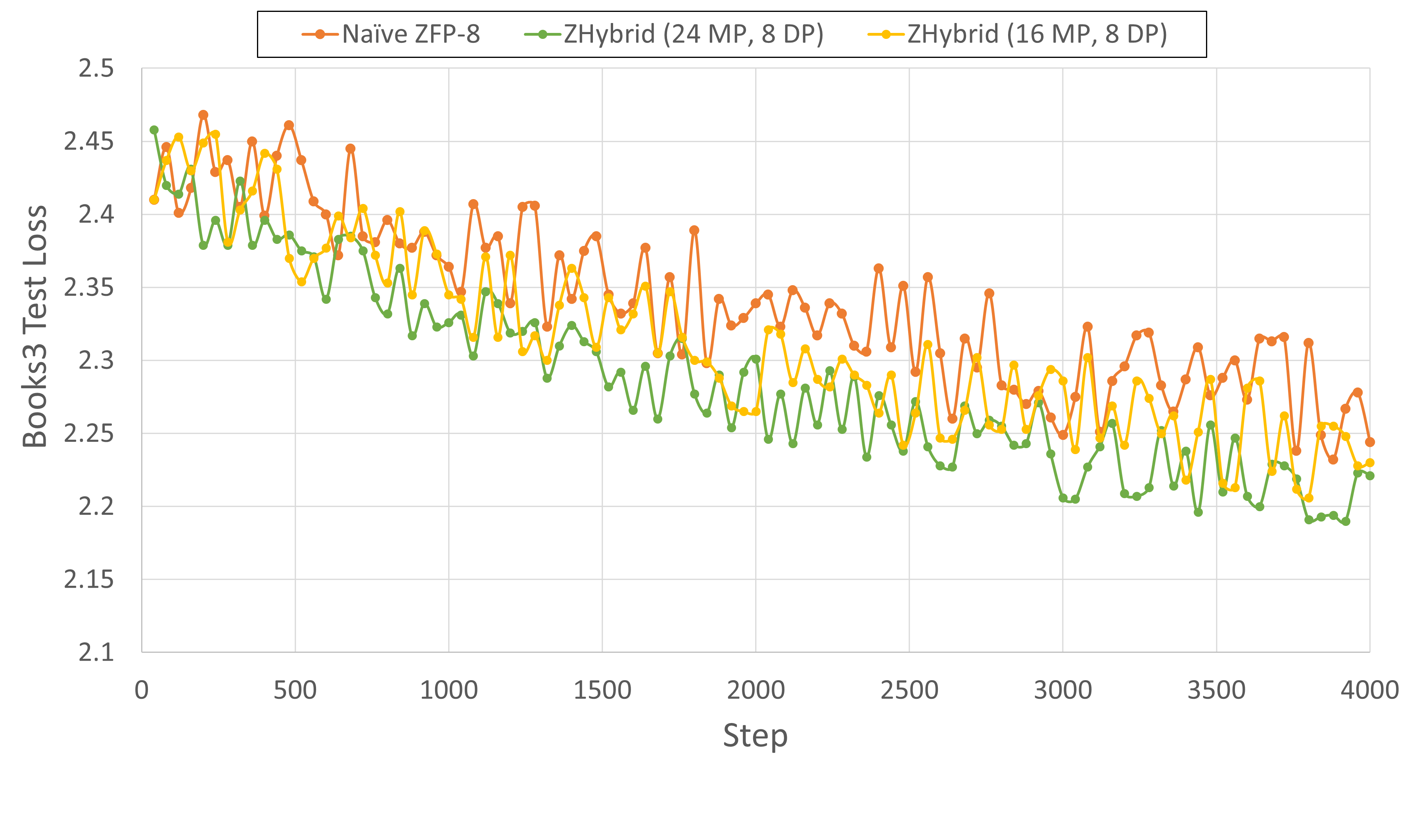}
    \caption{Comparison between \textit{ZHybrid} (rate:24 MP \& rate:16 MP) and naïve ZFP solution in test loss.} 
\label{fig:zhybrid-compr}
\end{figure}

\subsection{Discussions}
\label{sec:discussions}

We compare our hybrid compression approach with NCCL \cite{nccl}, a collective communication library highly optimized for NVIDIA GPUs and networking. Our approach \textit{ZHybrid}(rate:16-MP, rate:8-DP) exhibits up to \textbf{7.6\%} increase in samples per second and \textbf{12.9\%} increase in TFLOPS per GPU on 192 V100 GPUs. While higher ZFP rate (rate:24-MP, rate:8-DP) results in less performance gain, we still achieved up to \textbf{4.9\%} increase in samples per second and \textbf{5.5\%} increase in TFLOPS per GPU on the same scale.(Figure 12, 13) The reason for this is that as we scale up, inter-node bandwidth begins to saturate. Compression-assisted MPI collectives are capable to reduce message size during transfer and mitigate communication stress in this scenario, resulting in better GPU compute utilization. Compared to \textit{ZHybrid}, \textit{MZHybrid} provided more benefits on loss convergence rather than training throughput due to the overhead of lossless compression. The two proposed hybrid schemes benefit us in either training efficiency or quality; specific choices depend on the end-user's preference and metrics.

\begin{figure}[htbp]
\centering
    \includegraphics[width=0.7\columnwidth]{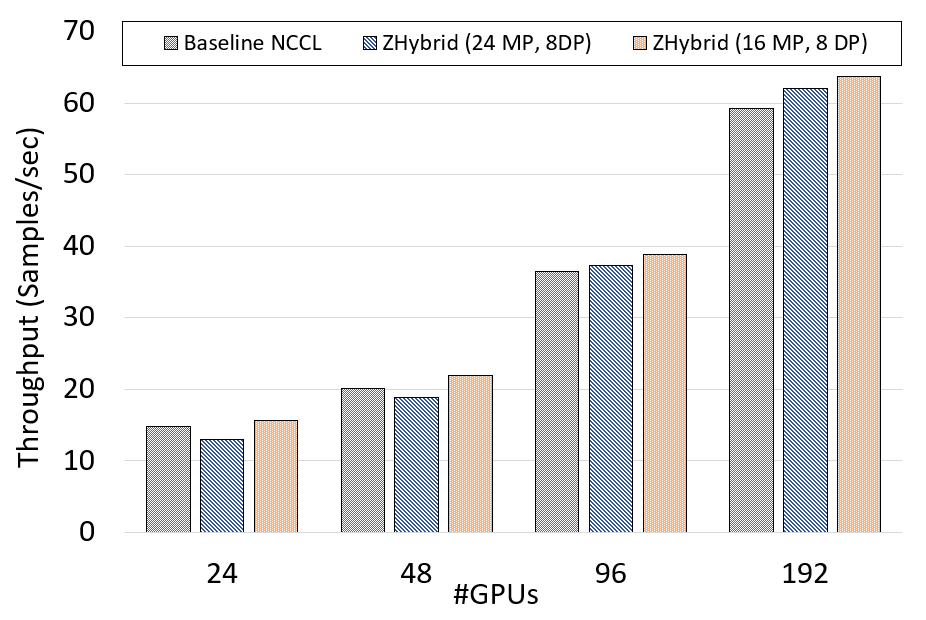}
    \caption{Comparison between \textit{ZHybrid} and NCCL in samples per sec.} 
\label{fig:zhybrid-nccl-samples}
\end{figure}

\begin{figure}[htbp]
\centering
    \includegraphics[width=0.7\columnwidth]{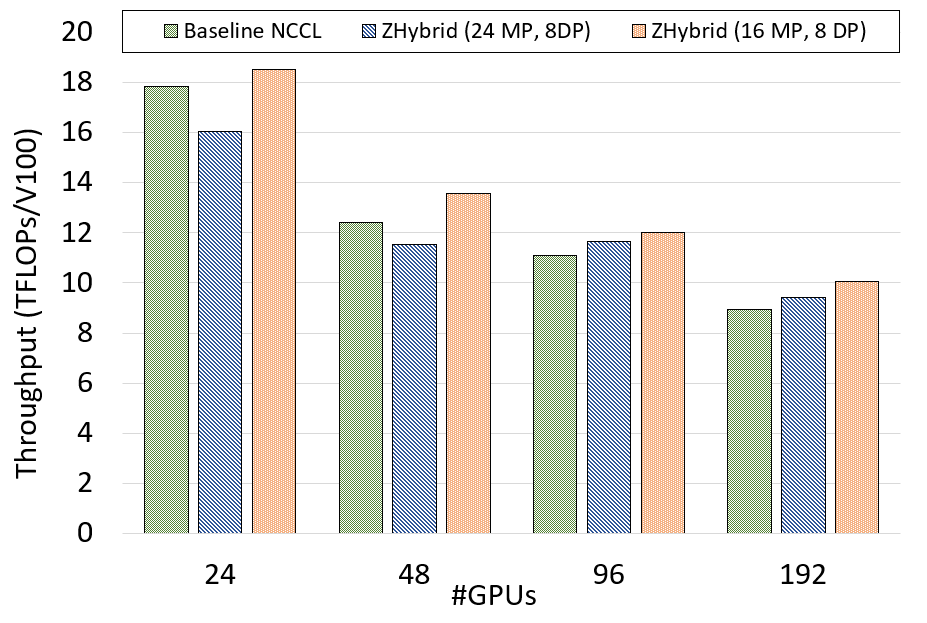}
    \caption{Comparison between \textit{ZHybrid} and NCCL in TFLOPS per GPU.} 
\label{fig:zhybrid-nccl-tflops}
\end{figure}




The key takeaway we addressed is that higher ZFP compression rates(i.e., less aggressive compression)lead to loss closer to baseline than low ZFP rates, which matches intuition. There is no correct answer for a “proper” rate since this depends on the use case, and this paper seeks to quantify the ZFP rate tradeoff so that this “proper” rate can be selected. The broad guidelines we find are to choose a model-parallel(TP+PP) compression rate slightly smaller than an FP-32 precision but large enough to provide a loss increase that our application can tolerate. At this stage, there is no detailed analysis for every rate, and all that is left to future work.


\section{Related work}
\label{sec:relatedworks}


Several studies have been done on compressing gradients to reduce training time \cite{signSGD, Deep-Grad-Compression}. Furthermore, C. C. Chen et al. \cite{chen-grad-compression} provided a hybrid communication compression method, which chooses the best compression method for every gradient to maintain model accuracy while reducing training time. Several distributed Deep Learning frameworka have also been integrated with mixed-precision training \cite{micikevicius2018mixed}.

These studies primarily focus on gradient compression for DP, with limited work dedicated to compression techniques in MP. Current efforts have also been attributed to mostly compressing gradients during training but hardly on activations. GPU-based compression has also been co-designed with various MPI collectives to speed up distributed DL training under a certain degree of parallelism, such as PyTorch-FSDP \cite{compression-redscat-allgather}.

Similar hybrid communication schemes has also been featured in works like MCR-DL \cite{mcr-dl}. MCR-DL supports mix-and-matching communication backends across MPI and NCCL \cite{nccl} for all point-to-point and collective operations. However, this work is not based on MCR-DL. The implementation invokes compression-assisted MPI collectives directly on the PyTorch level. The only reason MCR-DL is mentioned is to refer to its communication profiling insights \cite{mcr-dl}.

\section{Conclusion}
\label{sec:conclusion}
In this paper, we propose two hybrid compression schemes, namely \textit{MZHybrid} and \textit{ZHybrid}, to leverage LLM training efficiency with adequate model performance. These two designs consider the fundamental differences among parallelism strategies (DP, PP, and TP) and the sparsity in the message communicated in these schemes. \textit{MZHybrid} applied lossless MPC towards model-parallel communication and lossy ZFP towards data-parallel communication. \textit{ZHybrid} forces different ZFP rates for different parallelism stages. For model-parallel, we chose high-rate ZFP to preserve the precision of dataflow occurring within a model. For data parallel, we adopt low-rate ZFP to reduce gradient size communicated across models.

The proposed design \textit{MZHybrid} demonstrates up to 4.4\% increase in training samples per second and 5.3\% increase in TFLOPS per GPU compared to non-compression approaches with significant improvement in training quality over na\"ive compression. The proposed design \textit{ZHybrid} demonstrates up to 20.4\% increase in training samples per second and 20.6\% increase in TFLOPS per GPU compared to non-compression approaches and still poses noticeable improvement in model performance over na\"ive compression—the two hybrid schemes emphasized on training throughput benefits or model quality. 

We believe our approach can generalize to other use-cases which exhibit heavy collective communication and are saturating interconnect bandwidth.

In future work, we plan to co-design more MPI collective operations with GPU-based compression libraries to accelerate more scientific applications and Deep Learning workloads. Choices include All-reduce and other communication routines. Also, we expect that more advanced compression techniques and libraries can be incorporated such as cuSZ \cite{cuSZ} and block-based quantization \cite{dettmers20228bit}.

\bibliographystyle{IEEEtran}
\bibliography{bibfiles/lang.bib}
\eject
\end{document}